# WET: Weighted Ensemble Transformer – A Model for Identifying Psychiatric Stressors Related to Suicide on X (Formerly Twitter)

Ali Sahandi[1]*, Mahsa Pahlavan Yousefkhani[2], Mehrshad Eisaei[3], Hossein Momeni[4], Ramin Mousa[5]

[1] Department of Literature and Humanities, Shahid Beheshti University, Tehran, Iran, (sahandiali.sci@gmail.com)
[2] Department of Human Development and Family Science, Syracuse University, USA, (mpahlava@syr.edu)
[3] Department of Computer Science, University of Mazandaran, Babolsar, Iran, (mehhrshadeisaei017@gmail.com)
[4] Department of Computer Engineering, Golestan University, Gorgan, Iran, (h.momeni@gu.ac.ir)
[5] Department of Computer Science and Technology, University of Zanjan, Iran,( raminmousa@znu.ac.ir)

**Corresponding author:**
Ali Sahandi
Email: sahandiali.sci@gmail.com
ORCID: (0009-0004-0232-0072)

**Abstract**

Suicide remains one of the leading causes of death worldwide, particularly among young people, and psychological stressors are consistently identified as proximal drivers of suicidal ideation and behavior. In recent years, social media platforms such as X have become critical environments where individuals openly disclose emotional distress and conditions associated with suicidality, creating new opportunities for early detection and intervention. Existing approaches, however, predominantly rely on raw textual content and often neglect auxiliary emotional and contextual signals embedded in user metadata. To address this limitation, we propose a Weighted Ensemble Transformer (WET), a dual-branch deep learning architecture designed to identify psychiatric stressors associated with suicide in X posts. Our model integrates semantic representations extracted through Transformer encoders with an engineered feature vector capturing sentiment, subjectivity, polarity, and user engagement characteristics. We collected, filtered, and annotated 125,754 English tweets for suicide-related psychological stressors and evaluated the proposed model under two configurations. Extensive comparative experiments against traditional machine learning methods, advanced recurrent networks, and transformer baselines demonstrate that WET achieves state-of-the-art performance, reaching 0.9901 accuracy in binary classification.

**Keywords:** Psychiatric stressors, Suicide detection, Social media analytics, Transformer models, Computational mental health

## 1 introduction

Suicide persists as a critical and complex global public health challenge. According to the World Health Organization (WHO), over 703,000 people die by suicide annually, establishing it as a leading cause of death, particularly among young people (World Health Organization 2019). While global prevention initiatives have made progress, this trend is alarmingly reversing in some parts of the world. In the United States, for instance, suicide deaths reached an unprecedented peak in 2022, the highest number ever recorded and a rate unsurpassed since 1941 (Stone 2023). The causes of suicide and suicidal behaviors can be complex, varying significantly across individuals and societies. However, leading psychological theories of suicide converge on the critical role of psychiatric stressors. These stressors are identified as the acute triggers that can activate underlying vulnerabilities (Zalsman et al. 2016) and as the primary source of the intense psychological pain, feelings of entrapment, and perceived burdensomeness that fuel suicidal ideation (Van Orden et al. 2010; Klonsky and May 2015; O'Connor and Kirtley 2018). Therefore, the ability to identify these stressors is fundamental to understanding the pathway to suicide and developing appropriate and accurate intervention strategies.

Individuals increasingly rely on social networking platforms as essential features of their daily routines. The open environment of these networks provides a powerful medium for people to articulate their thoughts, ideas, and perspectives, thus serving as a major information source (Brailovskaia et al. 2019). Notably, these platforms have seen a rise in the open expression of suicidal thoughts. The combination of potential anonymity and the inherent openness of social media has lowered inhibitions, enabling users to candidly share their most negative feelings and emotional states (Mbarek et al. 2022). Twitter has been recommended by many scholars as a reliable source of information to analyze people's mental health conditions (O'dea et al. 2015; Luo et al. 2020). Monitoring these platforms can serve as an efficient approach for early risk detection, enabling potential interventions for at-risk users.

The evolution of Natural Language Processing (NLP) has provided powerful tools for this task, marked by a paradigm shift from traditional methods to sophisticated deep learning models. Although foundational transformer architectures like BERT have set high standards for capturing semantic context, these models are typically applied to raw text alone. This approach often overlooks valuable auxiliary information, such as sentiment and user engagement metrics, which provide crucial context for interpreting social media content related to mental health.

To address this gap, this paper introduces a novel approach for extracting psychiatric stressors from Twitter, centered on three primary contributions:

- A Novel Hybrid Architecture: We propose the Weighted Ensemble Transformer (WET), a dual-branch deep learning model designed to synergistically integrate deep semantic representations from a Transformer network with a curated vector of high-level features (e.g., sentiment, subjectivity, engagement metrics).
- A Large-Scale Annotated Dataset: We developed and annotated a new, large-scale dataset of over 120,000 tweets for the presence of psychiatric stressors related to suicide. This manually validated corpus serves as the foundation for our experiments and provides a significant resource for the research community.
- State-of-the-Art Performance: We conduct a comprehensive evaluation demonstrating that this architecture significantly outperforms a wide range of baselines, including traditional machine learning classifiers, recurrent neural networks, and standard transformer models.

## 2 Related Work

Significant research efforts have recently concentrated on detecting suicidality and identifying psychiatric disorders and suicide risk factors through the analysis of social media data. The convergence of artificial intelligence (AI) and mental health has driven significant advances in this area, particularly through the computational analysis of user-generated online content. Recent developments in natural language processing (NLP), machine learning (ML), and deep learning (DL) have enabled the development of increasingly sophisticated models capable of detecting indicators of suicidality. The following section reviews the related literature.

Several works in the field emphasize resource-efficient lexicon- and sentiment-based techniques. For instance, Sarsam (Sarsam et al. 2021) employed affective lexicons and semi-supervised learning to detect suicide-related tweets, demonstrating that emotional indicators such as sadness and fear are strong predictors of suicidal ideation. Similarly, Kumar & Venkatram (Kumar and Venkatram 2024) developed a rule-based decision tree model to assess suicidal risk levels using Twitter data. By employing a quicksort-based technique to identify optimal split points, their model achieved high classification accuracy, exceeding 90% across most node distributions. Manu Chandran Nair (Manu Chandran Nair 2024) analyzed over 25,000 depression-related tweets using topic modeling and sentiment analysis, identifying themes such as suicidal ideation, negative life experiences, and hopelessness. Lim & Loo (Lim and Loo 2023) developed a multiclass classification framework using a combination of TF-IDF, sentiment analysis (VADER), and part-of-speech tagging to distinguish between low, medium, and high suicide risk in tweets. Their model, trained with Random Forest, achieved over 86% accuracy and emphasized the importance of contextual linguistic features, showing that words like "kill" and "die" were more predictive of high risk, while PoS tags had limited impact. Bruinier et al. (Bruinier et al. 2024) focused on binary classification and demonstrated that incorporating sentiment features significantly improved model performance. Their Random Forest model, trained on 9,120 tweets, achieved 95.1% accuracy and 95.09% F1-score. Joarder et al. (Joarder et al. 2024) focused on predictive accuracy by testing multiple supervised machine learning models on a labeled dataset distinguishing between depression and suicide watch. Among the models, CatBoost achieved the highest accuracy (73.2%), followed closely by LightGBM. Deep learning and transformer-based models are increasingly utilized by researchers for suicide risk detection. Mirtaheri et al. (Mirtaheri et al. 2024) proposed a hybrid AL-BTCN model combining bidirectional temporal convolutional networks, LSTMs, and self-attention mechanisms. Their architecture achieved over 94% accuracy in detecting suicidal ideation from Reddit and Twitter posts. Similarly, Schoene et al. (Schoene et al. 2022) advanced the field by introducing a graph convolutional neural network (GCN) that integrated emotional and linguistic features, capturing fine-grained variations in emotional dominance and suicidality. Metzler et al. (Metzler et al. 2022) analyzed 3,202 tweets to explore how social media content relates to suicidal behavior. They categorized tweets into 12 themes, including suicidal ideation, recovery, awareness, and unrelated content. Using both traditional (SVM with TF-IDF) and deep learning models (BERT, XLNet), they conducted multi-class and binary classification tasks. Deep learning models generally outperformed SVM, with BERT achieving the best results in binary classification. Pokrywka et al. (Pokrywka et al. 2024) evaluated advanced transformer models for multi-class suicide risk detection, classifying posts into nuanced categories such as ideation, behavior, and attempt. Their fine-tuned GPT-4o model achieved top performance in the IEEE Big Data Cup. Holmes et al. (Holmes et al. 2025) provided a broader scoping review that confirmed BERT and GPT architectures as the most frequently used LLMs in suicidality research, while also highlighting the underrepresentation of ethical discourse in AI applications.

Other studies emphasize localization, language diversity, and socio-technical integration. Abdulsalam et al. (Abdulsalam et al. 2024) introduced the first Arabic suicidality detection model using AraBERT, achieving 91% accuracy. Likewise, Benjachairat et al.(Benjachairat et al. 2024) applied the Columbia-Suicide Severity Rating Scale (C-SSRS) in a Thai-language Twitter corpus, linking severity-level detection to a web-based cognitive behavioral therapy (CBT) platform. Metzler et al. (Metzler et al. 2024) extended suicide content classification to broadcast media using transformer models, contributing to large-scale surveillance of harmful and protective messaging based on established suicide reporting guidelines. Bello et al. (Bello et al. 2023) followed a similar approach, training CNN models on suicide-tagged tweets to detect themes in news articles, thereby bridging social and traditional media modalities. Emerging innovations also include fusion models and novel architectures. Dadgostarnia et al. (Dadgostarnia et al. 2025) applied capsule networks and IndRNNs to identify psychiatric stressors in Persian tweets. This approach achieved a binary classification accuracy of 83% in distinguishing tweets expressing suicide-related stressors from unrelated content, while Álvarez-López & Castro-Sanchez (Álvarez-López and Castro-Sánchez 2024) demonstrated that GPT-3 embeddings with deep neural networks surpassed traditional word embedding methods in classification performance. Ensemble learning approaches, as seen in Sharma & Neema (Sharma and Neema 2023) have also proven effective, particularly when balancing precision and recall in real-time applications. As summarized in Table 1, prior work spans lexicon/feature-based methods and transformer architectures, yet few studies combine deep contextual representations with engineered affective and behavioral signals

Table 1 Summary of related works

| Authors | Year | Features | Techniques/Methods | Data Source/Dataset | Category |
|---|---|---|---|---|---|
| Mirtaheri et al. Mirtaheri et al. (2024) | 2024 | Temporal/contextual features in text | Bi-TCN, LSTM, Self-Attention | Twitter, Reddit | Computer-based |
| Sarsam et al. (2021) | 2021 | Emotional and sentiment features | Lexicon-based model (YATSI), NRC, SentiStrength | Twitter | Computer-based |
| Metzler et al. (2024) | 2024 | Protective/harmful suicide-related features in media | TF-IDF+SVM, BERT | Broadcast media transcripts | Computer-based |
| Holmes et al. (2025) | 2025 | LLM applications in suicide prevention | Scoping review of transformer models | Multiple sources (PubMed, IEEE, etc.) | Review |

| Lim & Loo (2023) & Loo | 2023 | Multi-class suicide risk levels | TF-IDF, POS, VADER, Random Forest | Twitter | Computer-based |
|---|---|---|---|---|---|
| Benjachairat et al. (2024) | 2024 | Suicidal ideation severity (C-SSRS framework) | LSTM, RF, SVM, XGBoost | Thai Twitter | Computer-based |
| Schoene et al. (2022) | 2023 | Suicidal emotion classification | Graph Convolutional Network (GCN) | Twitter (TWISCO corpus) | Computer-based |
| Metzler et al. (2022) | 2022 | Harmful/protective tweet classification | BERT, XLNet | Twitter | Computer-based |
| Abdulsalam et al. (2024) | 2024 | Arabic suicidality | SVM (n-grams), AraBERT | Arabic Twitter | Computer-based |
| Pokrywka et al. (2024) | 2024 | Suicide risk level classification | DeBERTa, GPT-4o (fine-tuned) | IEEE BigData Cup | Computer-based |
| Bruinier et al. (2024) | 2024 | Sentiment-enhanced suicide detection | SVM, RF, LR, SGD, MNB | Twitter | Computer-based |
| Kumar & Venkatram (2024) | 2023 | Demographic & behavioral features | Rule-based decision tree | Twitter | Computer-based |
| Abdulsalam et al. (2024) | 2024 | ML model survey on social media | Systematic review | Twitter, Reddit | Review |
| Sharma & Neema (2023) | 2023 | Early intervention via ensemble models | Bagging, GradientBoosting | Twitter | Computer-based |
| Bello et al. (2023) | 2023 | Media theme detection | CNN | News articles (trained on Twitter data) | Computer-based |
| Dadgostarnia et al. (2025) | 2025 | Psychiatric stressor detection (Persian) | CapsuleNet, IndRNN | Persian Twitter | Computer-based |
| Manu Chandran Nair (2024) | 2024 | Emotional content and topic modeling | VADER, SentiArt, LDA | Twitter | Computer-based |
| Joarder et al. (2024) | 2024 | Depression/suicide risk | CatBoost, LightGBM, RF, ANN | Text data (unspecified) | Computer-based |
| Álvarez-López & Castro-Sanchez (2024) | 2024 | Embedding comparison | Word2Vec, GloVe, MPNet, GPT-3 + DNN | Social media | Computer-based |

## 3 Methodology

The proposed methodology for classifying tweets follows a structured, six stages pipeline, as illustrated in the Figure 1. This process is designed to systematically collect, preprocess, and analyze tweet data to train a sophisticated classification model. Each stage builds upon the previous one, starting from raw data collection and culminating in the application of a Transformer-based model. The following sections detail each component of this pipeline.

### 3.1 Collecting and Filtering Tweets

The initial step involved the creation of a dataset of potentially relevant tweets. To achieve this, we first manually compiled a comprehensive list of 90 keywords and phrases in English related to suicide and psychiatric stressors. The selection criterion for these terms was informed by previous research in the field, specifically the work of Du et al. (Du et al. 2018) and was augmented with clinical insights from a psychiatrist to include economic and social stimuli.

Using the Tweepy APIs, we collected public English tweets containing these keywords from 01, 2024 to 05, 2025. However, a manual review of this initial collection revealed that a significant portion of the tweets, while containing the target keywords, were not related to personal suicidal ideation. Many were news reports, promotional content, or discussions of terrorism (e.g., "suicide bomb", "suicide attack"), which introduced considerable noise and increased the potential for data misclassification and annotation burden.

### 3.2 Latent Feature Extraction

In this stage, we extract key textual features from the collected tweets using the TextBlob library. These features provide quantitative measures of the emotional and subjective content of the text, which serve as important inputs for our model. The extracted features include:

Subjectivity (s): Quantifies the degree of personal opinion versus factual information in the text.

Polarity (p): Measures the sentiment on a continuous scale from -1.0 (negative) to +1.0 (positive).

Sentiment (Se): Classifies the tweet into discrete categories of positive (1), neutral (0), or negative (-1).

In addition to these TextBlob features, we also extract metadata associated with each tweet, including Follower count (Fo), Likes (L), Replies (R), and Retweets (Re). Together, these extracted metrics form a "Feature vector" that provides rich, multi-modal information beyond the text itself. The output of this part is considered a latent feature.

### 3.3 Annotating Tweets

To prepare the data for training and evaluating our classification model, we annotated the filtered tweets with positive and negative tags. This process was crucial for creating a high-quality "golden tweet set" to train the deep learning model.

Positive Annotation: A tweet was labeled as positive if it expressed the user's personal experience, feelings, or thoughts related to suicide. This includes direct mentions of suicidal thoughts, plans, or a history of suicide attempts. For this task, the TextBlob library was utilized as an assistive tool to help identify content with relevant emotional and semantic indicators.

Negative Annotation: Tweets were labeled as negative if they met one of the following conditions, which were identified using manual rules:

1- Lack of connection: The tweet was entirely unrelated to the topic of suicide or suicidal thoughts.

2- Denial or rejection: The tweet explicitly denied suicidal thoughts or expressed hope, such as "I will not commit suicide."

3-Third-person discussion: The tweet discussed suicide or the suicidal thoughts of other people, rather than the user's own personal experience.

### 3.4 Preprocessing

Before feeding the textual data into the deep learning model, it undergoes a series of preprocessing steps to standardize and prepare it for analysis. This includes:

Stop Word Removal: To create a more accurate candidate set, we implemented a two-fold filtering strategy. First, during the collection process, we automatically excluded any tweets that contained URLs or specific phrases associated with non-personal events, such as "suicide attack." Second, we inductively generated a list of stop words based on a preliminary review of the data to further filter out irrelevant tweets. This rigorous filtering process resulted in a substantial increase in the relevance of the collected data. After applying these filtering steps, the final dataset consisted of 125,754 tweets, which served as the foundation for the subsequent annotation and modeling phases. Hashtags are handled, which may involve removal or treatment as regular text.

BERT Embedding: The cleaned text of each tweet is converted into a high dimensional numerical representation using a pre-trained BERT (Bidirectional Encoder Representations from Transformers) model. These embeddings capture deep contextual and semantic relationships between words, providing a powerful input for the classifier.

### 3.5 Train/Test Split

The annotated and preprocessed dataset is partitioned into training and testing sets. We employ a standard 80/20 split, where 80% of the data is used to train the model, and the remaining 20% is held out to evaluate its performance and generalization capability on unseen data.

### 3.6 Proposed Model

The final stage involves training and deploying our classification model. The architecture is built around a Transformer network, which is highly effective at understanding contextual nuances in sequential data like text. The model is designed to accept two primary inputs: the BERT embeddings generated from the tweet text and the auxiliary "Feature vector" extracted in Stage 2. A Weighted mechanism is incorporated into the model. The output of this model is the final classification for each tweet.

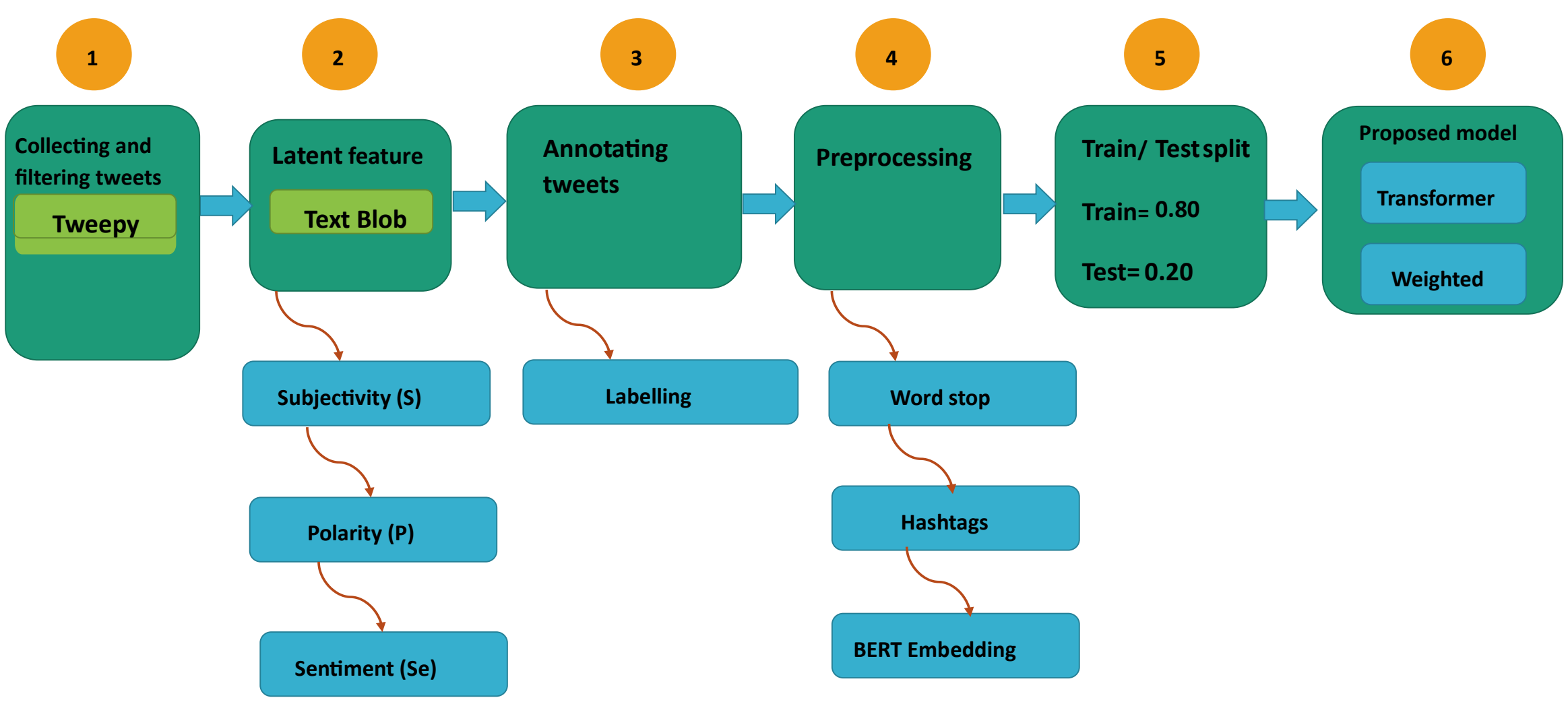

**Fig.1** The complete pipeline for Tweet data processing and classification, detailing the sequence from raw data ingestion, through data preprocessing and feature extraction, to the final model prediction

In a way, output combinations and model-based ensembles seek to improve learning models for different problems globally. Averaging will work better when differentiation among the various models exists; such differentiation might be through distinct hyperparameter settings or through differences in the data on which each member model is trained. This is because two neural networks with the same architecture and hyperparameter settings produce near-identical outputs; averaging their outputs is equivalent to selecting the output from any one of those models. In, Cireşan et al. (Ciregan et al. 2012)

MCDNN (Multi-Column Deep Neural Networks) created routes of data input, which were finally fitted to different models for each piece of data, with the result obtained by averaging model outputs. On the other hand, the procedure explained in Frazão et al. (Frazão and Alexandre 2014) held constant the inputs and the data preprocessing fed to each network and then applied weights at the output from the networks to mix and average the results. Inspired by such methodologies, we shall now explore our Weighted Ensemble Transformer (WET) in this section. The overall architecture of WET is illustrated in Figure 2, and the details are provided below.

A foundational challenge in applying Transformers to text is their inherent permutation-invariance. To understand sequence order, they rely on injected positional encodings. The original Transformer used a sinusoidal absolute position encoding. For a position $i$ and embedding dimension $k$, the formula is:

$$p_i^{(2k)} = \sin(i.w_k) \quad , \quad p_i^{(2k+1)} = \cos(i.w_k) \tag{1}$$

However, this can suffer from a loss of distance awareness and anisotropic vector distribution, degrading performance on nuanced tasks. To address this, our Transformer blocks incorporate two advanced positional encoding strategies: Foumani et al. (Foumani et al. 2024) token Absolute Position Encoding (tAPE) and efficient Relative Position Encoding (eRPE).

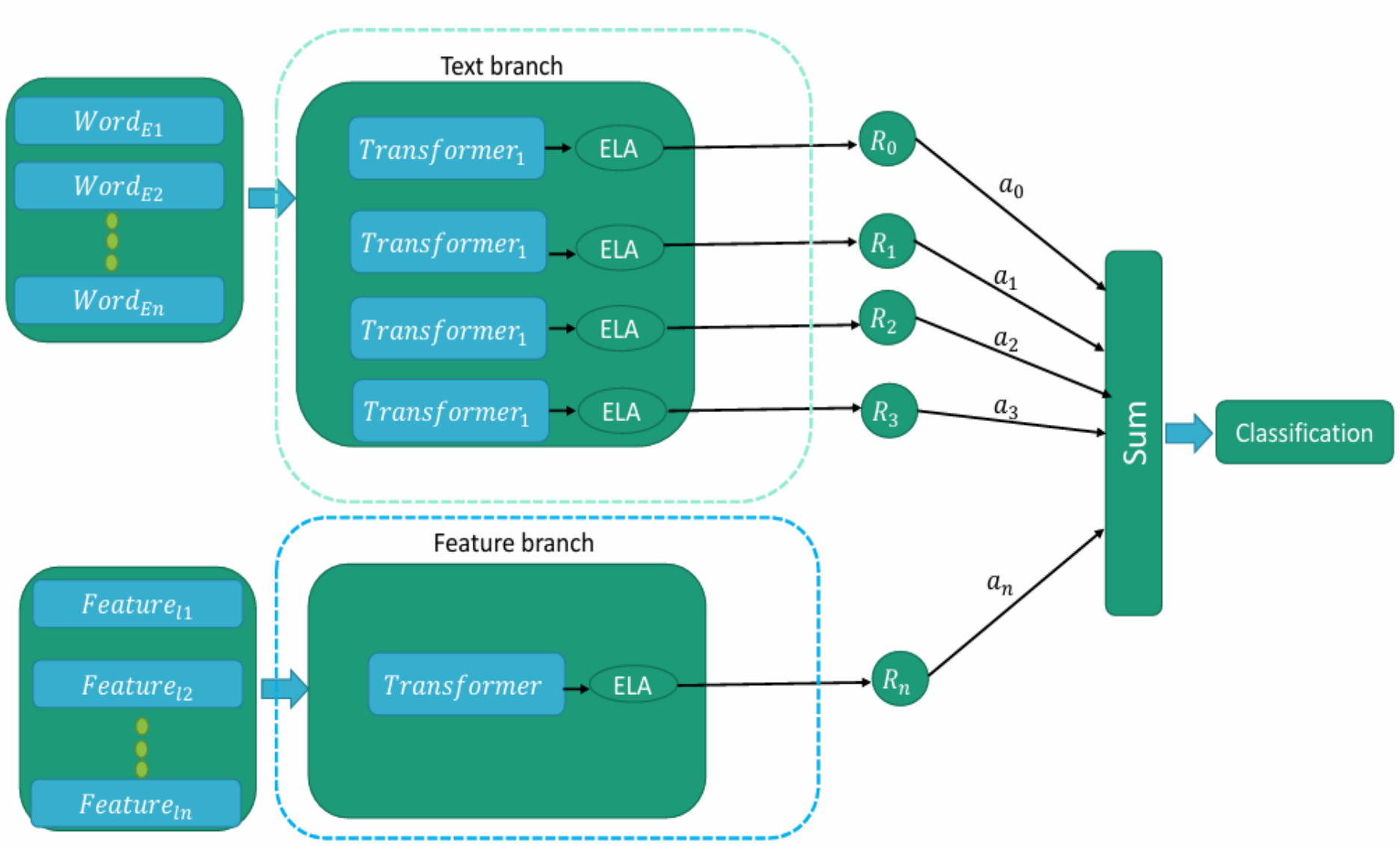

**Fig.2** Architecture of the Weighted Ensemble Transformer (WET) model. The proposed dual-branch architecture consists of: (1) a text branch (left) that processes BERT word embeddings through multiple parallel Transformer layers enhanced with Efficient Local Attention (ELA) modules to generate diverse semantic representations ($R_0$-$R_n$), and (2) a feature branch (right) that processes auxiliary features (sentiment, polarity, engagement metrics) through a Transformer-ELA combination. The outputs from both branches are combined using learned weights ($a_0$-$a_n$) and aggregated through summation to produce the final classification for detecting suicide-related psychiatric stressors in social media posts. The ELA modules enable precise attention to salient features while the weighted ensemble mechanism allows dynamic prioritization of different representational perspectives

### 3.6.1 Text branch: Deep Semantic Analysis

The Text Branch serves as the model's primary engine for semantic understanding, processing the BERT embeddings generated in Stage 4. This branch features an internal ensemble of four parallel Transformer layers, each producing a distinct representation ($R_0$, $R_1$, $R_2$, $R_3$). This parallel structure allows the model to capture diverse and complementary contextual patterns from the same input text. Each Transformer block within this branch is enhanced with the following mechanisms:

To provide a robust and stable signal of the absolute token position, we replace the standard sinusoidal encoding with tAPE. This method adapts the encoding formula to be sensitive to both the sequence length ($L$) and the embedding dimension ($d_{model}$), ensuring that the resulting vectors are both distance-sensitive and isotropic. The frequency term $\omega_k$ in the standard sinusoidal formula is updated to $\omega_k^{new}$ as follows:

$$w_k^{new} = \frac{w_k \, . d_{model}}{L} \quad where \quad w_k = \frac{1}{1000^{2k/d_{model}}} \tag{2}$$

The resulting tAPE vectors are added to the input BERT embeddings, providing a superior positional foundation for the self-attention mechanism. To explicitly model the crucial pairwise distances between tokens, we integrate eRPE. Rather than modifying the query and key vectors, eRPE introduces a trainable scalar bias directly into the attention calculation. This allows the model to efficiently learn the importance of relative positioning (e.g., the influence of a negation word on a sentiment word three positions away). The final attention output $\alpha_i$ for a token $i$ is computed as:

$$\alpha_i = \sum_j \left( softmax\left( \frac{Q_i K_j^T}{\sqrt{d_k}} \right) + w_{i-j} \right) . V_j \tag{3}$$

Where:

The standard attention score between query $i$ and key $j$ is:

$$e_{i,j} = \frac{Q_i K_j^T}{\sqrt{d_k}} \tag{4}$$

The softmax attention weight is obtained as:

$$\alpha_{i,j} = \frac{exp(e_{i,j})}{\sum_{k \epsilon L} exp(e_{i,k})} \tag{5}$$

.

$w_{i-j}$ is the learnable bias indexed by the relative distance between tokens $i$ and $j$, allowing the model to integrate relative positional context directly into the value aggregation step.

Finally, $V_j$ denotes the input (value embedding) at position $j$.

**3.6.2 Efficient Local Attention (ELA)**

ELA functions as a computational unit that enables deep networks to accurately select regions of interest or key object locations (Xu and Wan 2024). The core of ELA is Coordinate Attention (CA). CA consists of two parts: coordinate information embedding and coordinate attention generation. During the embedding of coordinate information, the authors consider a smart way to capture long-range spatial dependencies through strip pooling. Considering a convolutional block $R^{H*W*C}$, the strip pooling does average pooling (AVGpooling) over each channel in two spatial ranges $H$ and $W$ respectively such that one output representation is derived for the c-th channel at height $h$ and another output representation is derived for the c-th channel at width $w$ Xu & Wan(Xu and Wan 2024). The exact process can be seen mathematically as follows:

$$z_c^h(h) = \frac{1}{H} \sum 0 \leq i < H x_c \ (h, i) \tag{6}$$

$$z_c^w(w) = \frac{1}{w} \sum 0 \leq j < W x_c \ (j, w) \tag{7}$$

Now, the resulting equations are combined to generate the feature map that is passed on to the joint transformation function $F_1$ and $BN$. This is expressed as follows Xu &(Xu and Wan 2024):

$$f = \delta( BN(F_1 ([z^h , \quad z^w]))) \tag{8}$$

In this equation, $[-; -]$ and $\delta$ denote the cascade operation along the spatial dimension and the nonlinear activation function.

Suppose now that $f^h \in R^{(c/r*H)}$ and $f^w \in R^{(c/r*H)}$ belong to the spatial dimension. Then $g_c^h$ and $g_c^w$ would be these two 1∗1 convolutional transformations obtained from Xu &Wan(Xu and Wan 2024):

$$g_c^h = \sigma\left(F_h\left(f^h\right)\right) \tag{9}$$

$$g_c^w = \sigma\left(F_w\left(f^w\right)\right) \tag{10}$$

Finally, CA module can be represented as *Y:*

$$y_c(i,j) = x_c(i,j) \, . \, g_c^h(i) \, . \, g_c^w(j) \tag{11}$$

Normalization CA can be modified as follows:

$$y^h = \sigma\left(G_n\left(F_h(z_h)\right)\right) \tag{12}$$

*y* represents the ELA.

### 3.6.3 Feature branch

The Feature Branch is designed to process the auxiliary "Feature vector" extracted in Stage 2. This vector provides crucial, high-level context that is often not explicitly available in the raw text, containing features such as subjectivity (s), polarity (p), sentiment (Se), follower count (Fo), likes (L), and retweets (Re).

The process of converting a raw time series dataset into a supervised learning format for classification involves several key and fundamental steps. The raw data is divided into sections, each identified by $X_i$ where *i* denotes the section index (in our dataset, *i* refers to a record). Each section $X_i$ . contains a subset of the original data and is associated with a specific label $y_i$ . These sections are defined by their shape, which consists of *S* channels and *W* samples. Here, *S* corresponds to the number of channels in the data and *W* represents the number of samples within each section. First, a delayed input sequence is created that reflects the past observations leading up to the current time step. This lagged sequence, denoted $X_{i,lagged}$ , is generated by adding data from the same segment that has been shifted by a specified number of time steps. This process is repeated for a range of lag values, determined by the parameter $n_{in}$. The resulting form of $X_{i,lagged}$ is $(S,\ W \cdot n_{in})$.

After the lagged input sequence is created, a forecast sequence is constructed for each segment. This forecast sequence, denoted $X_{i,forecast}$, represents future observations after the current time step. Similar to the lagged sequence, $X_{i,forecast}$ is generated by adding data from the segment that has been shifted forward in time. The number of shifts is determined by the parameter $n_{out}$ .

The form of $X_{i,forecast}$ is $(S,\ W \cdot n_{out})$.

Once the delayed input sequence and the forecast sequence are prepared, they are horizontally concatenated to form the transformed part $X_{i\,,transformed}$. This new sequence contains both past and future information and has the form $(S,\ W \cdot (n_{\text{in}} + n_{\text{out}}))$.

This process is repeated for all records, resulting in a set of transformed parts $\{X_{i\,,transformed}, y_i\}$ for each $i$ . Finally, the transformed supervised dataset

*D* is created, which contains tuples $\{X_{i\,,transformed}, y_i\}$ for each segment *i* .

To apply the Transformer to this data, first, we map the scalar field $x_{ti}$ to the model-dim vector *d* with an LSTM network. Second, we use fixed-position embedding to represent the local texture, which can be represented as follows:

$$PE_{(pos,2j)} = \sin\left(\frac{pos}{2L_x}\right)^{2j/d_{model}} \tag{13}$$

$$PE_{(pos,2j+1)} = \cos\left(\frac{pos}{2L_x}\right)^{2j/d_{model}} \tag{14}$$

The i-m query attention can be defined as a kernel in the following possible form:

$$\text{Attention } (q_i, K, V) = \sum \frac{k(q_{i}, k_j)}{\sum k(q_{i}, k_l)} v_j \tag{15}$$

The dispersion measure of query $i$ can be defined as follows. This measure is used to identify the most important queries:

$$M(q_i, K) = \ln\left(\sum_{j=1}^{l_k} exp\left(\frac{q_i K^T}{\sqrt{2}}\right)\right) - \frac{1}{L_k} \sum_{j=1}^{l_k} exp\left(\frac{q_i k^T}{\sqrt{2}}\right) \tag{16}$$

ProbSparse Self-attention can be defined as:

$$Attention(Q, K, V) = softmax\left(\frac{\bar{Q} K^T}{\sqrt{d}}\right) V \tag{17}$$

The encoder module is used to capture the long-term dependence of the inputs. The t-th record $X_t$ is mapped to the matrix $X^t_{feed_{e}n} \epsilon R^{Lx} * d_{model}$. This module consists of several attention layers and LSTM layers. The procedure between the two layers is calculated as follows:

$$X^t_{j+1} = ELU(LSTM\ M([X^t_j]_{AB})) \tag{18}$$

Where $[.]_{AB}$ is the Multi-head ProbSparse self-attention.

To dynamically identify and prioritize the most predictive signals within this vector, this branch incorporates the Efficient Local Attention (ELA) module. The output of the ELA module—a re-weighted feature vector where salient features are amplified and irrelevant ones are suppressed—is then passed to a Transformer encoder layer. This layer learns the complex, non-linear interdependencies among the prioritized features, producing the final, context-aware feature representation, R.

## 4. Model Description

Given some input patterns, the output probabilities from all models are averaged before making a prediction. For output $i$, the average output $S_i$ is calculated as follows:

$$S_i = \frac{1}{n} \sum_{j=1}^{n} R_j(i) \tag{19}$$

Where $R_j(i)$ is the output $i$ of network $j$ for a given input pattern.

Our approach involves applying different weights to each network. In the validation set, networks that had lower classification errors will be given higher weights when combining the results. Given an input pattern, the output probabilities from all models are multiplied by a weight $a$ before prediction:

$$S_i = \sum_{j=1}^{n} a_j R_j(i) \tag{20}$$

## 5 Results

To evaluate the performance of our proposed WETransformer model, we conducted a comprehensive set of experiments comparing it against a diverse range of baseline and state-of-the-art models. The evaluation was performed on the task of extracting psychiatric stressors related to suicide from Twitter data. All models were assessed using the standard metrics of accuracy, precision, recall, and F1-score to ensure a holistic comparison.

$$Accuracy = \frac{(TP+TN)}{(TP+FN+TN+FP)} \tag{21}$$

$$Precision = \frac{(TP)}{(TP+FP)} \quad (22)$$

$$Recall = \frac{(TP)}{(TP+FN)} \quad (23)$$

$$F1-score = \frac{(2*precision*recall)}{(precision+recall)} \quad (24)$$

Where *TP* (True Positive) is the number of positive samples correctly classified as positive, *FP* (False Positive) is the number of negative samples incorrectly classified as positive, *TN* (True Negative) is the number of negative samples correctly classified as negative, and *FN* (False Negative) is the number of positive samples incorrectly classified as negative. For the evaluation, the dataset was partitioned into an 80% training set and a 20% test set. To ensure a comprehensive analysis, the performance of our proposed model was benchmarked against several baseline methods.

**5.1 Ablation study**

The ablation study is completed with different element variations in the proposed model, and the results are recorded. The time complexity and accuracy of the experiment are evaluated for each experimental configuration. The results of the ablation studies performed with the experimental (external) dataset are presented in Tables 2 and 3, where Table 2 summarizes all the results related to the configuration of the Dropout rate, activation function, Batch size, and fully connected, and Table 3 shows some of the results of parameter tuning and loss function.

**5.2 Ablation Study Results**

Dropout Rate: Dropout is a widely used regularization technique in deep neural networks to reduce overfitting. In simpler terms, it works by randomly dropping out a subset of neurons during training. When a neuron is dropped, it is excluded from both forward and backward propagation, and its incoming and outgoing connections are temporarily removed. Formally, each neuron is retained with probability $p$ (the keep rate) and dropped with probability $(1 - p)$. By preventing over-reliance on individual neurons, this random process promotes the learning of more robust and generalized feature representations. At test time, dropout is disabled and the full network is used, with weights appropriately scaled to reflect the dropout applied during training. The model was evaluated with dropout rates of 0.5, 0.6, 0.7, and 0.8, all while maintaining a constant time complexity of 75.45 M. The results indicate that a dropout rate of 0.5 achieved the highest test accuracy of 0.9925. Increasing the dropout rate led to a decline in performance, with rates of 0.7 and 0.8 yielding modest accuracies of 0.9922 and 0.9923 respectively, and a rate of 0.6 resulting in the lowest accuracy of 0.9921 among the tested variations.

Activation Function: The choice of activation function is critical for introducing non-linearity. Five functions were evaluated: Linear, PReLU, LeakyReLU, Tanh, and ELU. The LeakyReLU function yielded a test accuracy of 0.9925, which served as the benchmark for this case. The Linear function performed notably poorly with an accuracy of 0.9612, while PReLU (0.9916), Tanh (0.9901), and ELU (0.9903) also resulted in lower accuracies.

Time complexity remained constant at 75.45 M across all tests.

Batch Size: Batch size refers to the number of training samples processed simultaneously before the network's parameters are updated. The investigation into batch size revealed a direct correlation between batch size and computational cost. As the batch size increased from 8 to 128, the time complexity rose from 75.45 M to 79.48 M. The highest test accuracy of 0.9925 was achieved with the smallest batch size of 8. Larger batch sizes of 32, 64, and 128 did not improve performance, yielding accuracies of 0.9905, 0.9919, and 0.9910, respectively.

Fully Connected Layer Size: To test the effect of increasing model capacity, the number of neurons in the fully connected layer was varied (64, 128, 256, 512). Similar to batch size, increasing the number of neurons in the fully connected layer from 64 to 512 resulted in a significant increase in time complexity from 75.45 M to 98.48 M. The configuration with 64 neurons provided the highest test accuracy of 0.9925. Larger layer sizes did not lead to improved accuracy; in fact, a slight performance degradation was observed, with accuracies of 0.9924 for 128 and 256 neurons, and 0.9917 for 512 neurons.

Optimizer: Four different optimizers were compared, with the time complexity held constant at 75.45 M. The Adam optimizer demonstrated superior performance, achieving a test accuracy of 0.9925. The other tested optimizers, including Nadam (0.9812), SGD (0.9819), and RMSprop (0.9910), all resulted in lower accuracy scores.

Loss Function: The final experiment evaluated five different loss functions to determine the most suitable objective for the model. The Binary Crossentropy loss function achieved the highest test accuracy of 0.9925. Other functions, such as Categorical Crossentropy (0.9893), Mean squared logarithmic error (0.9892), Mean Squared Error (0.9855), and Mean absolute error (0.9854), all failed to surpass this performance benchmark. The time complexity for all configurations in this case was 75.45 M.

the ablation studies reveal that no single variation improved upon the baseline model's performance. The configuration yielding the highest test accuracy of 0.9925 consistently emerged across all case studies as the optimal choice. This optimal configuration consists of a dropout rate of 0.5, the LeakyReLU activation function, a batch size of 8, a 64-neuron fully connected layer, the Adam optimizer, and the Binary Crossentropy loss function. Notably, the experiments demonstrate that increasing model complexity through larger batch sizes or a larger fully connected layer increases computational time without providing any corresponding benefit to test accuracy.

Table 2 Ablation study regarding changing Dropout rate, activation function, Batch size, and fully connected layer size.

| Case study 1: Dropout rate | Number of blocks | Time complexity (Million) | Test accuracy | Finding |
|---|---|---|---|---|
| 1 | 0.5 | 75.45 M | 0.9925 | Highest accuracy |
| 2 | 0.6 | 75.45 M | 0.9921 | Modest accuracy |
| 3 | 0.7 | 75.45 M | 0.9922 | Modest accuracy |
| 4 | 0.8 | 75.45 M | 0.9923 | Lowest accuracy |
| Case study 2: changing activation function | Activation Function | Time complexity (Million) | Test accuracy | Finding |
| 1 | Linear | 75.45 M | 0.9612 | Not Improved accuracy |
| 2 | PRelue | 75.45 M | 0.9916 | Not Improved accuracy |
| 3 | LeakyReLu | 75.45 M | 0.9925 | Not Improved accuracy |
| 4 | Tanh | 75.45 M | 0.9901 | Not Improved accuracy |
| 5 | ELU | 75.45 M | 0.9903 | Not Improved accuracy |
| Case study 3: Batch size | Size of batch | Time complexity (Million) | Test accuracy | Finding |
| 1 | 8 | 75.45 M | 0.9925 | Not Improved accuracy |
| 2 | 32 | 76.98 M | 0.9905 | Not Improved accuracy |
| 3 | 64 | 78.04 M | 0.9919 | Not Improved accuracy |
| 4 | 128 | 79.48 M | 0.9910 | Not Improved accuracy |
| Case study 4: Fully connected size | Number of neuron | Time complexity (Million) | Test accuracy | Finding |
| 1 | 64 | 75.45 M | 0.9925 | Not Improved accuracy |
| 2 | 128 | 78.98 M | 0.9924 | Not Improved accuracy |
| 3 | 256 | 80.04 M | 0.9924 | Not Improved accuracy |
| 4 | 512 | 98.48 M | 0.9917 | Not Improved accuracy |

Table 3 Ablation study regarding changing optimizer and Loss Function

| Case study 5: changing optimizer | Optimizer type | Time complexity (Million) | Test accuracy | Finding |
|---|---|---|---|---|
| 1 | Adam | 75.45 M | 0.9925 | Not Improved accuracy |
| 2 | Nadam | 75.45 M | 0.9812 | Not Improved accuracy |
| 3 | SGD | 75.45 M | 0.9819 | Not Improved accuracy |
| 4 | RMSprop | 75.45 M | 0.9910 | Not Improved accuracy |
| Case study 6: changing Loss Function | Loss function | Time complexity (Million) | Test accuracy | Finding |
| 1 | Binary Crossentropy | 75.45 M | 0.9925 | Not Improved accuracy |
| 2 | Mean Squared Error | 75.45 M | 0.9855 | Not Improved accuracy |
| 3 | Mean absolute error | 75.45 M | 0.9854 | Not Improved accuracy |
| 4 | Mean squared logarithmic error | 75.45 M | 0.9892 | Not Improved accuracy |
| 5 | Categorical Crossentropy | 75.45 M | 0.9893 | Not Improved accuracy |

**Performance of Traditional Machine Learning Baselines**

We first established a performance benchmark using traditional machine learning classifiers: Logistic Regression (LR), Support Vector Machine (SVM), Multinomial Naive Bayes (NB), Random Forest (RF), and Gradient Boosting (GB), including ensembled and Adaboost variations. These models were evaluated using three distinct n-gram feature sets: uni-gram, bi-gram, and a combination of both (see Table 4). The results demonstrated a clear dependency on the richness of the input features. While uni-gram models like Adaboost achieved an F1-score of 90.00%, performance degraded significantly when using bi-grams alone, where Ensembled was the top performer with an F1-score of 79.81%. The most effective configuration for traditional models was the combination of unigram and bi-gram features. In this setup, the Support Vector Machine (SVM) emerged as the strongest traditional baseline, achieving an accuracy of 92.92% and an F1-score of 92.78%. To determine if the performance differences among the models were statistically significant, a series of t-tests were conducted. (see Table 5)

Table 4 Summary of Performing Traditional Machine Learning Models by N-gram Type

| Model | N-gram | Accuracy | Precision | Recall | F1 |
|---|---|---|---|---|---|
| Logistic regression (LR) | Uni-gram | 88.983051 | 88.983051 | 88.983051 | 87.568841 |
| Support Vector Machine (SVM) | | 89.830508 | 89.830508 | 89.830508 | 88.761048 |
| Naive Bayes (Multinomial) | | 88.644068 | 88.644068 | 88.644068 | 87.405814 |
| Random Forest(RF) | | 88.474576 | 88.474576 | 88.474576 | 86.838825 |
| Gradient Boosting Classifier(GB) | | 88.983051 | 88.983051 | 88.983051 | 87.632025 |
| Ensembled | | 90.000000 | 90.000000 | 90.000000 | 88.986234 |
| Adaboost | | 90.000000 | 90.000000 | 88.829308 | 90.000000 |
| Logistic regression (LR) | Bi-gram | 79.322034 | 79.322034 | 79.322034 | 74.145486 |
| Support Vector Machine (SVM) | | 81.864407 | 81.864407 | 81.864407 | 78.581050 |
| Naive Bayes (Multinomial) | | 81.355932 | 81.355932 | 81.355932 | 77.354883 |
| Random Forest(RF) | | 74.237288 | 74.237288 | 74.237288 | 65.727039 |
| Gradient Boosting Classifier(GB) | | 80.169492 | 80.169492 | 80.169492 | 75.701448 |
| Ensembled | | 82.542373 | 82.542373 | 82.542373 | 79.811369 |
| Adaboost | | 74.915254 | 74.915254 | 74.915254 | 66.924242 |
| Logistic regression (LR) | Uni-gram and Bi-gram | 92.000000 | 92.000000 | 92.000000 | 91.789819 |
| Support Vector Machine (SVM) | | 92.923077 | 92.923077 | 92.923077 | 92.784749 |
| Naive Bayes (Multinomial) | | 91.692308 | 91.692308 | 91.692308 | 91.537193 |
| Random Forest(RF) | | 87.384615 | 87.384615 | 87.384615 | 86.807061 |
| Gradient Boosting Classifier(GB) | | 89.692308 | 89.692308 | 89.692308 | 89.495368 |
| Ensembled | | 91.846154 | 91.846154 | 91.846154 | 91.690365 |
| Adaboost | | 90.615385 | 90.615385 | 90.615385 | 90.343847 |

Table 5 T-test for traditional machine learning models by N-gram Type

| | LR | SVM | NB | RF | GB | Ensembled | Adaboost |
|---|---|---|---|---|---|---|---|
| Logistic regression (LR) | 1.0 | 0.00071 | 0.00735 | 0.34725 | 1.80364 | 0.00017 | 0.002232 |
| Support Vector Machine (SVM) | 0.00071 | 1.0 | 0.131016 | 0.045519 | 0.65136 | 0.60535 | 0.89850 |
| Naive Bayes (Multinomial) | 0.00735 | 0.13101 | 1.0 | 0.279454 | 0.014265 | 0.03979 | 0.16469 |
| Random Forest(RF) | 0.34725 | 0.04551 | 0.27945 | 1.0 | 0.012160 | 0.01836 | 0.05593 |
| Gradient Boosting Classifier(GB) | 1.80364 | 0.65136 | 0.01426 | 0.01216 | 1.0 | 0.85431 | 0.82892 |
| Ensembled | 0.00017 | 0.60535 | 0.03979 | 0.01836 | 0.85431 | 1.0 | 0.75359 |
| Adaboost | 0.00223 | 0.89850 | 0.16469 | 0.05593 | 0.82892 | 0.75359 | 1.0 |

### 5.3 Performance of Deep Learning Baselines

We next evaluated a suite of deep learning architectures, ranging from recurrent networks to advanced transformer and capsule-based models, to establish a more competitive set of baselines (see Table 6). Recurrent-based models demonstrated a notable improvement over the n-gram approaches. The initial set of baselines consisted of models based on recurrent neural networks (RNNs) and a hybrid of a CNN and an RNN. The 1-layer LSTM and 2-layer LSTM both achieved an identical accuracy of 0.9195. However, the 1-layer LSTM demonstrated slightly superior Precision (0.9293), Recall (0.9299) and F1 (0.9295) compared to the 2-layer version (Precision: 0.9123, Recall: 0.9112, F1: 0.9117). This suggests that for this dataset, increasing the depth of the LSTM network did not yield a performance benefit and may have slightly hindered its ability to generalize. The GRU model performed comparably to the LSTMs, with an accuracy of 0.9146. Its Precision (0.9233) and Recall (0.9232) were indicating a level of performance equivalent to that of the LSTM models. The hybrid CNNLSTM model showed a notable improvement over the pure recurrent architectures, achieving an accuracy of 0.9392. It also recorded the highest Precision (0.9344), Recall (0.9345) and F1 (0.9344) within this group.

Table 6 Performance of Advanced Deep Learning Baselines

| Baseline | Accuracy | Precision | Recall | F1 |
|---|---|---|---|---|
| LSTM 1-layer | 0.9195 | 0.9293 | 0.9299 | 0.9295 |
| LSTM 2-layer | 0.9195 | 0.9123 | 0.9112 | 0.9117 |
| GRU | 0.9146 | 0.9233 | 0.9232 | 0.9232 |
| CNNLSTM | 0.9392 | 0.9344 | 0.9345 | 0.9344 |
| BERT | 0.9600 | 0.9600 | 0.9600 | 0.9600 |

| XLNet | 0.9700 | 0.9700 | 0.9700 | 0.9700 |
|---|---|---|---|---|

Significant performance improvement was observed with the introduction of transformer-based architectures. These models demonstrated a clear superiority over the recurrent and hybrid baselines. The Bert model achieved an exceptional and perfectly balanced performance, recording 0.9600 for Accuracy, Precision, Recall, and F1-score. This represents a substantial leap of approximately 2-3 percentage points in accuracy over the best-performing hybrid model (CNNLSTM). The XLNet model set the highest benchmark among all evaluated baselines. It surpassed Bert's performance, achieving 0.9700 across all four metrics: Accuracy, Precision, Recall, and F1-score.

The evaluation clearly delineates a performance hierarchy among the baseline models. The transformer architectures (XLNet and Bert) significantly outperform the recurrent and hybrid models. XLNet stands out as the top-performing baseline, establishing a very high standard for the classification task with an F1-score of 0.97. To provide a more challenging comparison, we also benchmarked against a range of highly specialized deep learning architectures. This included various

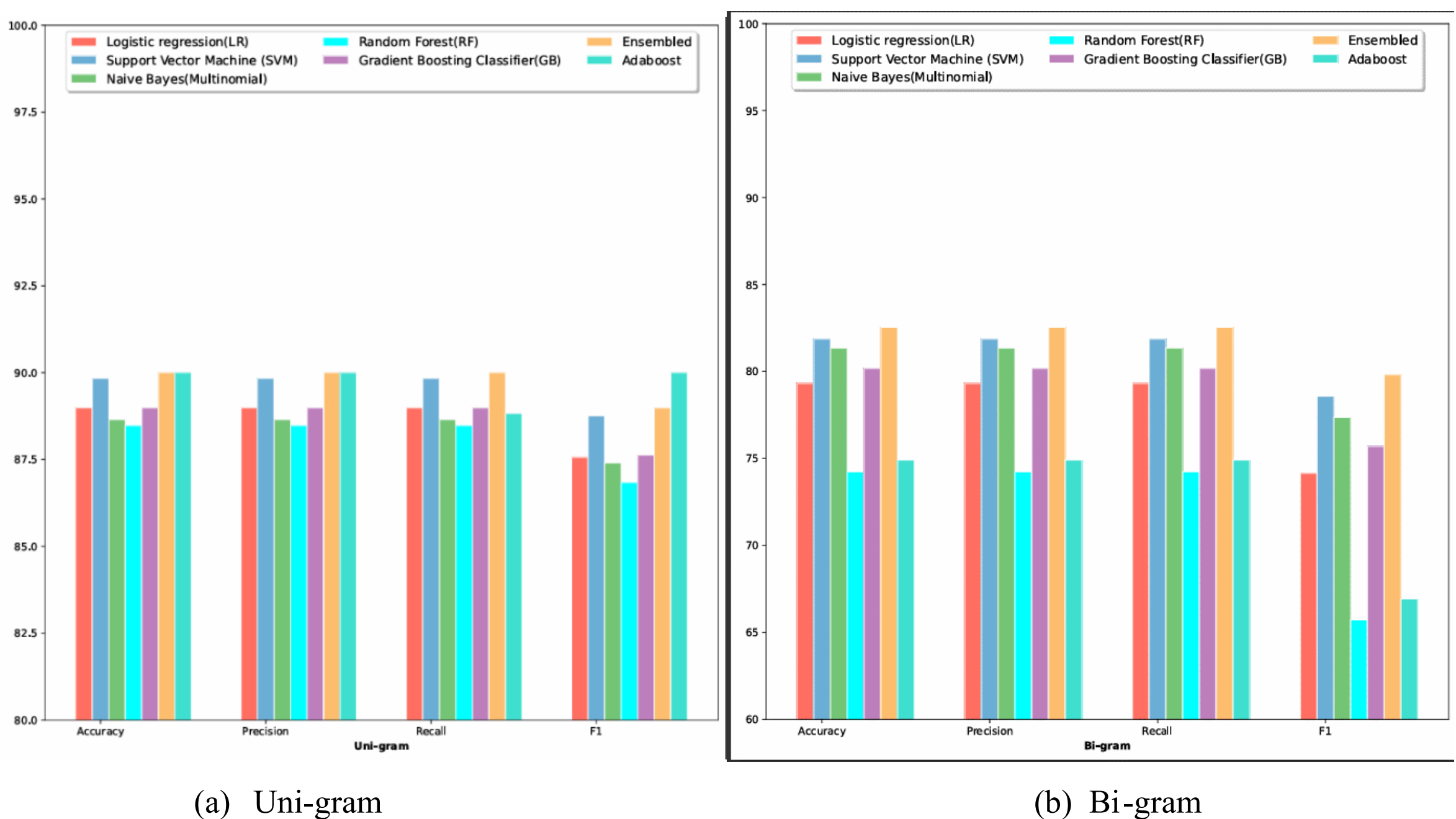

(a) Uni-gram (b) Bi-gram

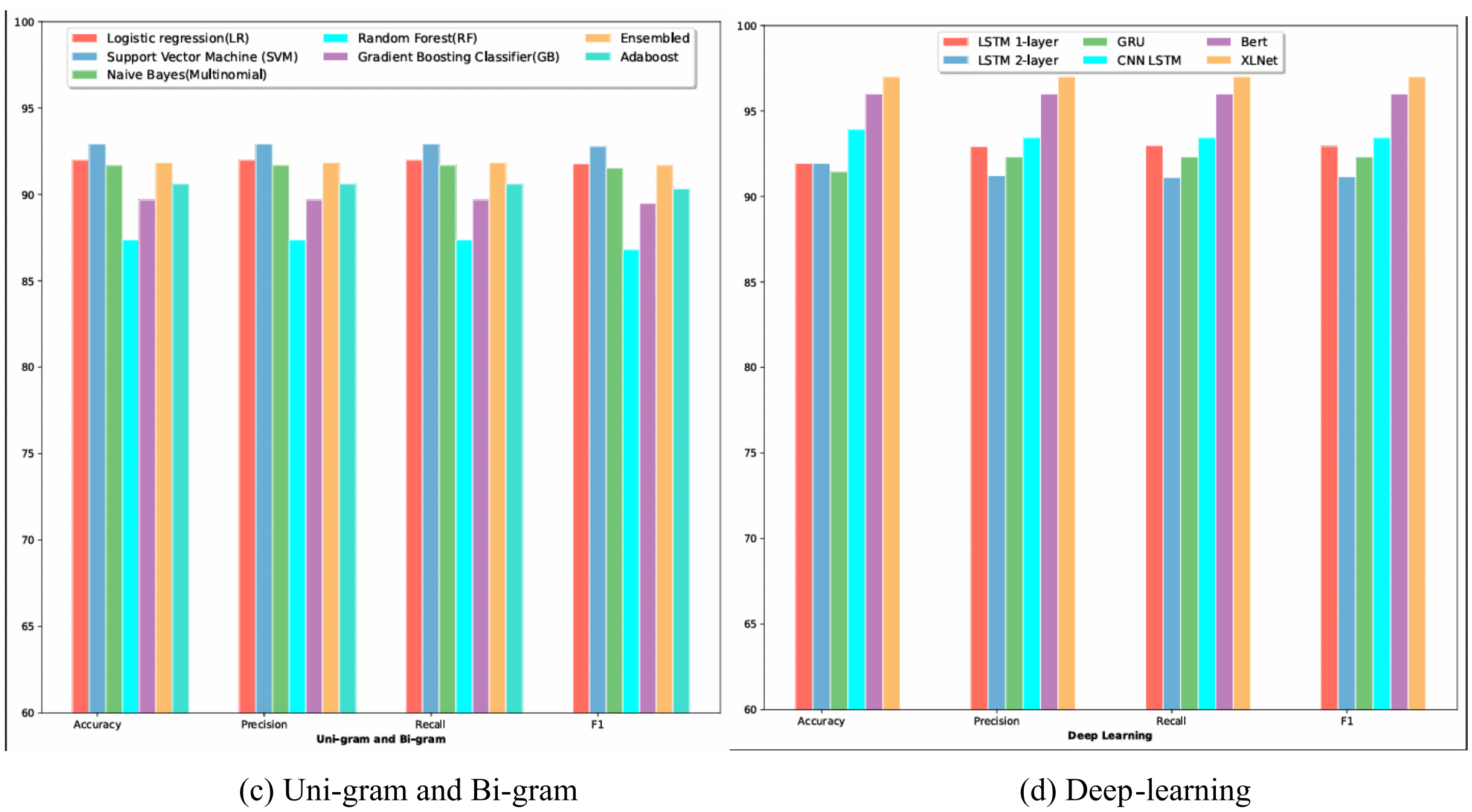

(c) Uni-gram and Bi-gram

(d) Deep-learning

**Fig.3** Performance comparison of classification models across different feature configurations: (a) traditional ML models using uni-gram features, (b) traditional ML models using bi-gram features, (c) traditional ML models using combined uni-gram and bi-gram features, and(d) deep learning models, showing accuracy, precision, recall, and F1- score metrics for suicide stressor detection.

CNNs, recurrent variants, as well as The Capsule Network Models In methods based on convolutional networks, several architectures were evaluated. For the Char-level CNN small model, an overall accuracy of 0.9342 was achieved. In the Positive class, this approach reached a Precision of 0.93, Recall of 0.94, and F1-score of 0.93. For the Negative class, it obtained a Precision of 0.94, Recall of 0.93, and F1-score of 0.94. The Char-level CNN large model obtained an accuracy of 0.9493. It achieved a Precision of 0.95, Recall of 0.95, and F1-score of 0.95 in the Positive class, and identical metrics of 0.95 for Precision, Recall, and F1 in the Negative class. The VDCNN-29 layers model reached an accuracy of 0.9555. This model obtained a Precision, Recall, and F1-score of 0.96 for the Positive class and also achieved a Precision, Recall, and F1-score of 0.96 for the Negative class. The Word-level CNN model achieved an accuracy of 0.9524. For the Positive class, it reached a Precision of 0.95, Recall of 0.96, and F1-score of 0.95. In the Negative class, it obtained a Precision of 0.96, Recall of 0.95, and F1-score of 0.95. Finally, the fast Text model achieved an accuracy of 0.9487. This approach reached a Precision of 0.96, Recall of 0.94, and F1-score of 0.95 in the Positive class. For the Negative class, it obtained a Precision of 0.94, Recall of 0.96, and F1-score of 0.95.

For recurrent neural network models, the D-LSTM achieved an accuracy of 0.9276. In the Positive class, it obtained a Precision of 0.91, Recall of 0.94, and F1-score of 0.93. For the Negative class, the model reached a Precision of 0.94, Recall of 0.92, and F1-score of 0.93. The IndRNN model reached an accuracy of 0.9352. For the Positive class, it obtained a Precision of 0.93, Recall of 0.94, and F1-score of 0.93. In the Negative class, the metrics were a Precision of 0.94, Recall of 0.94, and an F1-score of 0.94.

More advanced Capsule Networks, designed to preserve hierarchical spatial features, demonstrated even stronger performance. The Bi-GRUCapsule model achieved an overall accuracy of 0.9676. For the Positive class, it reached a Precision of 0.96, Recall of 0.98, and F1-score of 0.97. In the Negative class, it obtained a Precision of 0.98, Recall of 0.96, and F1-score of 0.97. The Capsule Fusion model reached an accuracy of 0.9790. This approach obtained a Precision of 0.98, Recall of 0.97, and F1-score of 0.98 in the Positive class. For the Negative class, it reached a Precision of 0.97, Recall of 0.98, and an F1-score of 0.98. The Bi-IndRNN Caps model achieved an accuracy of 0.9792. For the Positive class, its performance was a Precision of 0.97, Recall of 0.99, and F1-score of 0.98. In the Negative class, it obtained a Precision of 0.99, Recall of 0.97, and an F1-score of 0.98.

The proposed model was evaluated in two configurations, both of which set a new state-of-the-art. The Proposed model (small) achieved a superior accuracy of 0.9889. In the Positive class, it obtained a Precision of 0.99, Recall of 0.99, and F1-score of 0.99. The performance was identical in the Negative class, with a Precision, Recall, and F1-score of 0.99. The Proposed model (large) further improved performance, reaching an accuracy of 0.9901. This model achieved a Precision of 0.99, Recall of 0.99, and an F1-score of 0.99 for both the Positive class and the Negative class, demonstrating exceptional and perfectly balanced performance. (see Table 7). To visually synthesize the performance of the various architectures, Figure 4 provides a comparative bar chart of the overall model accuracies and Figure 5 breaks down the key classification metrics of Precision, Recall, and F1-score for both the "Potential Suicide" and "Not Suicide" classes

Table 7 Class-Wise Performance of Advanced Deep Learning Architectures

| Model | | Class | Precision | Recall | F1 | Accuracy |
|---|---|---|---|---|---|---|
| CNN | Char-level CNN small | Not Suicide | 0.94 | 0.93 | 0.94 | 0.9342 |
| | | Potential Suicide | 0.93 | 0.94 | 0.93 | |
| | Char-level CNN large | Not Suicide | 0.95 | 0.95 | 0.95 | 0.9493 |
| | | Potential Suicide | 0.95 | 0.95 | 0.95 | |
| | VDCNN–29 layers | Not Suicide | 0.96 | 0.96 | 0.96 | 0.9555 |
| | | Potential Suicide | 0.96 | 0.96 | 0.96 | |
| | Word-level CNN | Not Suicide | 0.96 | 0.95 | 0.95 | 0.9524 |
| | | Potential Suicide | 0.95 | 0.96 | 0.95 | |
| | fastText | Not Suicide | 0.94 | 0.96 | 0.95 | 0.9487 |
| | | Potential Suicide | 0.96 | 0.94 | 0.95 | |
| RNN | D-LSTM | Not Suicide | 0.94 | 0.92 | 0.93 | 0.9276 |
| | | Potential Suicide | 0.91 | 0.94 | 0.93 | |
| | IndRNN | Not Suicide | 0.94 | 0.94 | 0.94 | 0.9352 |
| | | Potential Suicide | 0.93 | 0.94 | 0.93 | |
| Capsule | Bi-GRUCapsule | Not Suicide | 0.98 | 0.96 | 0.97 | 0.9676 |
| | | Potential Suicide | 0.96 | 0.98 | 0.97 | |
| | Capsule Fusion | Not Suicide | 0.97 | 0.98 | 0.98 | 0.9790 |
| | | Potential Suicide | 0.98 | 0.97 | 0.98 | |
| | Bi-IndRNN Caps | Not Suicide | 0.99 | 0.97 | 0.98 | 0.9792 |
| | | Potential Suicide | 0.97 | 0.99 | 0.98 | |
| | | Not Suicide | 0.99 | 0.99 | 0.99 | |
| Proposed model | small | Potential Suicide | 0.99 | 0.99 | 0.99 | 0.9889 |
| | large | Not Suicide | 0.99 | 0.99 | 0.99 | 0.9901 |
| | | Potential Suicide | 0.99 | 0.99 | 0.99 | |

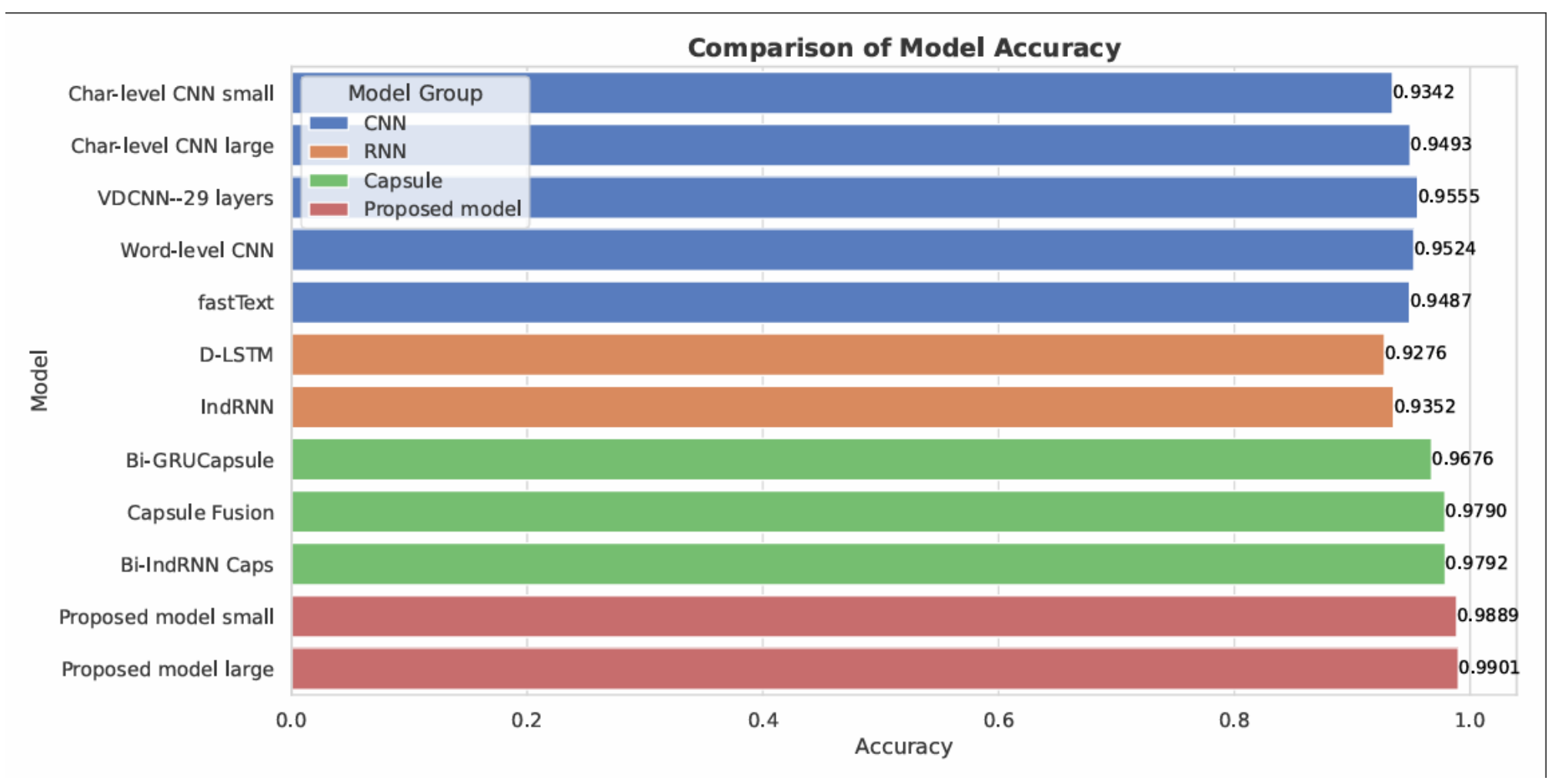

**Fig.4** Accuracy comparison of advanced deep learning architectures grouped by model type (CNN, RNN, Capsule, and Proposed WET model) for suicide stressor detection, with the proposed model achieving the highest accuracy of 99.01

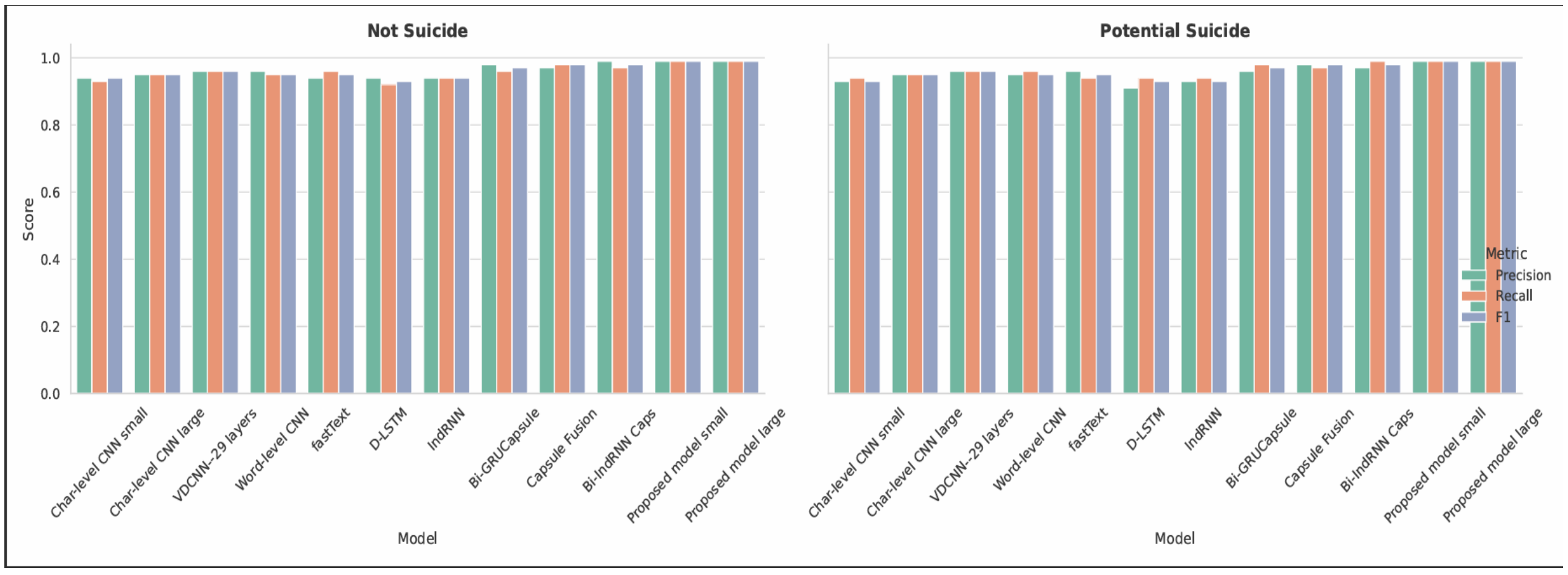

**Fig.5** Class-wise performance metrics (precision, recall, and F1-score) for advanced deep learning architectures on "Not Suicide" and "Potential Suicide" classification, demonstrating the proposed WET model's superior and balanced performance across both classes

## 6 Discussion

This study introduced the WETransformer, a novel deep learning architecture designed for the classification of psychiatric stressors in social media text. The experimental results demonstrate that the proposed model achieves a state-of-the-art accuracy of 99.01% and an F1-score of 99.00%, significantly outperforming a comprehensive suite of traditional machine learning, recurrent, and advanced transformer-based baselines. The performance hierarchy observed in our experiments where traditional n-gram models were surpassed by recurrent networks, which were in turn outperformed by standard transformer architectures like BERT and XLNet aligns with the general trajectory of the Natural Language Processing field (Garrido-Merchan et al. 2023). However, the WETransformer's ability to exceed the performance of even strong baselines like XLNet (97.00% F1-score) and specialized architectures like Bi-IndRNN Caps (98.00% F1-score) warrants closer analysis.

The key innovation of the WETransformer lies in its hybrid-input design, which synergistically combines deep semantic representations from BERT embeddings with a curated Feature vector of sentiment, subjectivity, and engagement metrics. This architecture implicitly validates the findings of prior studies that highlighted the predictive power of sentiment and emotional features (Sarsam et al. 2021; Bruinier et al. 2024) While those studies often incorporated such features into less complex models like Random Forest, our work demonstrates that fusing these explicit signals with the rich, contextual understanding of a transformer network yields a substantial performance gain. The model's success suggests that while BERT embeddings capture the nuanced semantics of the text, the auxiliary features provide complementary, high-level signals about the user's emotional state and the social resonance of their post, which are critical for disambiguating intent in the complex domain of mental health discourse.

Furthermore, the results contribute to the ongoing scholarly conversation about architectural enhancements beyond standard fine-tuning. Research on enhancing BERT architectures (Thakur et al. 2021; Warner et al. 2025) has shown the value of augmenting BERT with specialized components. Similarly, our approach can be seen as a form of architectural specialization, tailored specifically for social media analysis where both content and context are vital. The model's performance suggests that for this classification task, this fusion of semantic and engineered features approaches the ceiling of predictive accuracy on the given dataset.

### 6.1 Connection to Prior Work and Significance

Our findings both confirm and extend existing research in suicidality detection. The demonstrated superiority of transformer-based models over traditional methods is consistent with the conclusions of Metzler et al. (Metzler et al. 2022) and the broader review by Holmes et al. (Holmes et al. 2025) However, our study advances the field by proposing a concrete architectural improvement that pushes performance beyond what standard transformer models achieve. While many studies focus on optimizing a single modality (text), our work underscores the importance of a multi-modal approach, integrating linguistic data with readily available social and emotional metadata. This aligns with the socio-technical perspective that mental health expression online is not merely textual but is embedded in a system of social interaction.

The practical implications of this work are significant. A highly accurate and automated classification system like the WETransformer could serve as a powerful tool for public health surveillance, enabling researchers and organizations to monitor trends in psychiatric distress at a population level with unprecedented detail. This approach could complement large-scale efforts, such as those analyzing broadcast media (Metzler et al. 2024) by providing a real-time lens into public discourse. Moreover, it could form the technological backbone for proactive intervention systems, similar to the vision proposed by Benjachairat et al. (Benjachairat et al. 2024) who linked suicide risk detection to a web-based therapy platform. The reliability of our model could reduce the rate of false positives, which is a critical concern in deploying such systems to avoid unnecessary and potentially harmful outreach.

### 6.2 Limitations and Directions for Future Research

Despite the promising results, this study is subject to several limitations. First, our dataset, while large, is confined to English-language tweets. As demonstrated by research in Arabic and Thai, linguistic and cultural contexts are paramount in mental health expression. Therefore, the generalizability of our model to other languages and cultures remains untested.

Second, the data collection methodology, which relied on an initial set of 90 keywords, introduces a potential selection bias. This approach effectively captures explicit mentions of suicide and related stressors but may miss more implicit, coded, or novel expressions of distress. Consequently, our model is trained on a specific subset of online discourse and may not perform as well on data that deviates from these linguistic patterns.

Finally, it is crucial to emphasize that our model identifies correlations between linguistic patterns and labels of potential psychiatric distress; it does not and cannot perform clinical diagnosis or predict suicidal acts. The ethical application of this technology must be carefully considered to avoid overinterpretation of its outputs, a concern highlighted in the scoping review by Holmes et al. (Holmes et al. 2025) The findings and limitations of this study suggest several promising avenues for future investigation. A primary goal should be to validate the WETransformer architecture across different social media platforms (e.g., Reddit, which has distinct discourse norms) and in multiple languages to assess its robustness and adaptability. This would involve creating new, culturally-aware keyword lists and annotation guidelines.

Future work should also explore more sophisticated feature engineering for the auxiliary input vector. Beyond sentiment and basic engagement metrics, incorporating temporal features (e.g., time of day, posting frequency), network features (e.g., user connectivity), or psycholinguistic features derived from lexicons could provide even richer context for the model.

Finally, future research must address the ethical and practical challenges of deployment. This includes longitudinal studies to assess model performance over time and the development of human-in-the-loop systems that integrate this technology responsibly into mental health support workflows, ensuring that automated detection is a precursor to compassionate and effective human intervention.

## Statements and Declarations

**Competing Interests**:

The authors declare that they have no known competing financial interests or personal relationships that could have appeared to influence the work reported in this paper.

**Funding**:

The research received no specific grant

**Ethical Approval**:

This study analyzed English-language posts collected from the X platform (formerly Twitter) using the official API in accordance with the platform's Terms of Service and Developer Policy. Only publicly accessible posts from public user accounts were collected. No private messages, protected accounts, or restricted content were accessed.

Although the dataset generated for this study is not publicly shared, the original posts were publicly available at the time of data collection. The study did not involve direct interaction with individuals, intervention, or the collection of private identifiable information. Therefore, under standard institutional definitions, the research did not constitute human subjects research requiring formal ethics committee approval.

Given the sensitive nature of suicide-related content, strict data protection procedures were followed. Usernames, profile identifiers, and other directly identifying information were removed prior to analysis. Data were stored securely and used exclusively for academic research purposes. No attempt was made to identify, contact, or intervene with individual users.

The model developed in this study is intended solely for research and population-level analysis and does not provide clinical diagnosis or individual-level risk assessment.

**Data Availability**:

The dataset generated during the current study was constructed from posts collected via the X API and is subject to platform restrictions. Due to ethical considerations and compliance with X's Developer Policy, the raw tweet content cannot be publicly redistributed. Derived data or tweet identifiers may be made available from the corresponding author upon reasonable request and subject to platform and ethical compliance.